# Analysis of Voltage Stability in Terms of Interactions of $Q(U)$-Characteristic Control in Distribution Grids

Sebastian Krahmer [1)], Stefan Ecklebe [2)], Peter Schegner [1)], Klaus Röbenack [2)]
[1)] Chair of Electrical Energy Supply, Technische Universität Dresden, Dresden, Germany
[2)] Institute of Control Theory, Technische Universität Dresden, Dresden, Germany
sebastian.krahmer@tu-dresden.de

*Abstract*—As the amount of volatile, renewable energy sources in power distribution grids is increasing, the stability of the latter is a vital aspect for grid operators. Within the *STABEEL* project, the authors develop rules on how to parametrize the reactive power control of distributed energy resources to increase the performance while guaranteeing stability. The work focuses on distribution grids with a high penetration of distributed energy resources equipped with $Q(U)$-characteristic. This contribution is based on the stability assessment of previous work and introduces a new approach utilizing the circle criterion. With the aim of extending existing technical guidelines, stability assessment methods are applied to various distribution grids – including those from the SimBench project. Herein, distributed energy resources can be modelled as detailed control loops or as approximations, derived from technical guidelines.

## I. Introduction

Due to the increasing development of renewable energy sources in Europe (e.g. installation of around 105 GW of new wind energy capacity over the next five years [1]), grid operators are tasked with providing robust and reliable power distribution. In addition, the delivered energy of distributed energy resources (DERs) is not only subject to technical fluctuations due to the nature of their sources but also subject to economic fluctuations due to the increasing use in their direct marketing. These changes require not only an adaption of the existing electrical power grids but also their operation in terms of voltage level control or reactive power provisioning. Distribution grid operators (DSO) are entitled by the grid code [2] to request specific voltage support dependent on maximum power capability of the power plant and the voltage level of their grid connection point (PCC). The operators can meet these challenges by employing a combination of centralized and distributed voltage control concepts with regard to cost-effectiveness and efficiency, as shown by previous work [3, 4].

### A. Stability Assessment

The stability of voltage control can be evaluated in different time scales: starting with current control time scale as presented in work [5] and heading up to steady state voltage stability [6]. In the mid-range time scale work was presented regarding the automatic reactive power regulation as in [7] or central reactive power setpoint optimization as in [3]. However, possible interactions of $Q(U)$-characteristic controls are classified as (short-time) voltage stability or as slow-interaction converter-driven stability, newly introduced in [8]. Considering possible control interactions, grid codes are very restrictive regarding parameterization limits, which counteract a more grid-serving application and must therefore continue to be investigated. As [9] stated for DER dominated low-voltage distribution grids, there exists no limitations using a $Q(U)$-characteristic based control for photovoltaic (PV) following fundamental parameterization rules. These findings are based on an evaluation of the control overshoot. At higher voltage levels, single DERs are grouped together and equipped with farm control. Here, different voltage control strategies can be considered: centralized, distributed or decentralized [10, 11]. A communication delay within farm control must be taken into account as of this point. However, especially in weak grids the DSO must be aware of voltage control interactions, as shown in [12], and should know a general assessment method next to the recommendations of the grid codes, cf. [2]. A stability assessment based on wind farm (WF) models with focus on medium- and high-voltage levels is introduced in previous work [13]. This approach does not require knowledge of the model parameters, but results in conservative thresholds, i.e. remains in parts below the grid code recommendations. In this paper, the authors introduce a less conservative approach.

### B. Approach

In this work, an evaluation of the interaction of controllers is performed based on different DER models, as is depicted in Figure 1. Section II introduces DER models with $Q(U)$-characteristic, which can be expressed by detailed control loops or by approximated models in respect to technical guidelines (VDE-AR-N 4120 [14]). In Section III, a more convenient stability assessment approach based on the so-called circle criteria [15] is presented in addition to previous work [13]. An evaluation of different high-voltage distribution grids is carried out in Section IV. Thereby, both methods as well as numerical simulations in *PowerFactory* are be applied to these grids and the results are classified. Section V summarizes the conclusions.

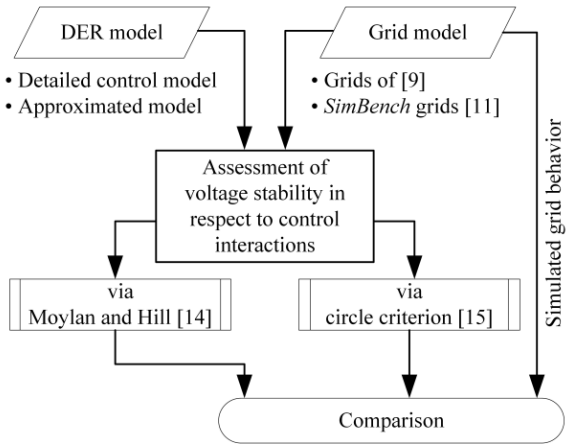

Figure 1. Proposed plan of evaluation.

## II. MODELS

In the first part of this section the authors summarize different DER models from the literature and unite them in a general detailed plant model. In the second part, two basic models are presented for cases where either model parameters or the detailed model structure is not known. The reactive power control is fed by a $Q(U)$-characteristic. Finally, a comparison is drawn on the example of a WF.

### A. Detailed Models of Voltage Control in Distributed Energy Resources (Orig. DER)

Normative publications and research work such as [9, 16, 17, 18] provide complex models of almost all different DER types. Assuming a normal operation mode and neglecting active power constraints the reactive power control path can be treated separately. Figure 2 shows a generic reactive power control path for a multiple-input multiple-output (MIMO) case adapted to the reactive power control via a $Q(U)$-characteristic, cf. appendix VI.A and VI.B. Herein, the reactive power control loop $G(s)$ of each plant can be separated from the nonlinear characteristic $\psi$ regardless of the fashion in which control is implemented inside of $G(s)$. The control interdependence of all DERs through the common grid model is mapped by the nodal sensitivity matrix $\boldsymbol{K}_Q$. Therefore, both linear blocks result in:

$$\widetilde{\boldsymbol{G}}(s) = \boldsymbol{G}(s) \cdot \boldsymbol{K}_Q. \qquad (1)$$

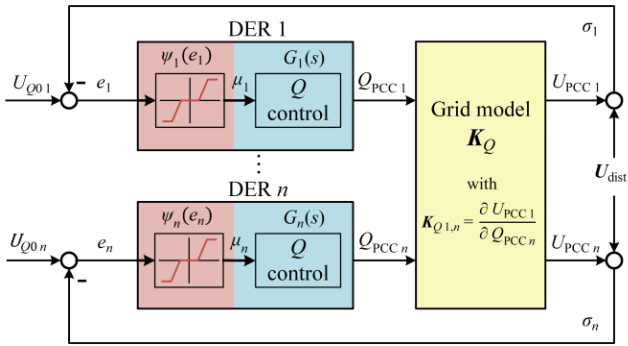

Figure 2. Generic reactive power control path for MIMO case, adapted to a control type of $Q(U)$-characteristic.

Thereby, $\boldsymbol{K}_Q$ is a constant $n \times n$ matrix that can be computed using the *Jacobian* obtained during the iterative solution of non-linear power flow equations, as presented in [13, 19]. A more realistic, i.e. a voltage dependent grid model is presented in [20], but not used, since it adds an additional nonlinearity.

Furthermore, the authors focus on three popular DER types: (i) WF from type fully rated converter (FRC), (ii) WF from type double fed induction generator (DFIG), (iii) photovoltaic farm (PVF). Detailed model configurations are proposed in the appendix TABLE III. Further models of reactive power resources like static var compensator (SVC) or static synchronous compensator (STATCOM) are not yet within focus, but can be found in [21, 22, 23]. Also, battery storages are not considered here due to their widely adaptable inverter behavior.

### B. Alternatives to the Detailed Models

If a detailed DER model is not available, two alternative approaches, named as PT2-DER and PT2-TAR, can be used. The first approach assumes that the DER operator does not supply a detailed model to the stability assessor, but provides a model approximation, e.g. as a PT2 element (2). This approximation can also be achieved by fitting a PT2 element to the frequency response of a fully parameterized, detailed DER model. The second approach can be interpreted as a fallback, if a detailed model or even an approximation based on this model is not available. Here, a PT2 element is fitted against a generic step response. Thus, it takes into account the principle dynamics of reactive power provision specified in grid codes / technical guidelines (TAR), e.g. in the German TAR [14].

#### 1) PT2 Fit Based on the Frequency Response of a Specific DER Model (PT2-DER)

Using the detailed DER model as fitting target, a realistic approximation can be provided. The drawback is the required specification of DER model parameters, which, however, are usually provided to the DSO when the DER is commissioned. For this approach, the authors are using the frequency response with respect to magnitude and angle within a relevant frequency band for PT2 fitting, cf. Figure 3 b. Herein, this frequency band describes the range within the control loop inputs are expected to lie. Inputs of other frequencies are filtered out in previous layers. The lower bound can be defined by the transformer tap control and the upper bound can be set at twice the voltage averaging time constant. Therefore, it follows $f_{\text{PT2 fit}} \in [10^{-2}, 10^2]$.

Furthermore, the PT2 gain $\kappa$ is fixed to 1 to match the static gain of the detailed model, cf. Figure 3 a. TABLE III shows the obtained PT2 parameters.

#### 2) PT2 Fit Based on Technical Guidelines (PT2-TAR)

When connecting DERs to distribution grids, technical guidelines apply, which regulate, e.g., the way in which reactive power is provided [2]. Thus, there exists specifications on the principle dynamics of reactive power provision. Incorporating these boundaries for a control response, a reduction of model complexity towards a PT2 representation is possible. The TAR high voltage [14] provides such a specification for the set of admissible control behaviors, by means of parameterization of a generic step response. The TAR specifies maximum overshoot $\xi$, rise time to first reach 90 % of static gain $T_{90\%}$ as well as settling time $T_{stl}$ to reach a tolerance band around the static gain. Assuming a "slow" parameterized DER, $\xi$ is set to 15 %, $T_{90\%} = 5$ s and $T_{stl}$ should around $T_{90\%} + 3$ s. At this point, a simple optimization algorithm based on least squares was used to fit a PT2 element of the form:

$$\text{PT2}(s) = \frac{\kappa}{1 + 2 \cdot D \cdot T \cdot s + T^2 \cdot s^2}. \tag{2}$$

Thereby the three characteristic parameters $\xi$, $T_{90\%}$ and $T_{stl}$ are faced with only two degrees of freedom $D$, $T$, assuming that the gain $\kappa$ is set to 1 to match the static gain of the step response. As shown in Figure 3, a match of $\xi$ and $T_{90\%}$ were weighted more significantly, because of the high impact on dynamic system response. TABLE III shows the obtained PT2 parameters.

### C. Comparison of Detailed and Approximated Models

Figure 3 shows the comparison of step and frequency response of a detailed DER model with its PT2 approximations. A WF-FRC model was used in this paper, cf. TABLE III (i).

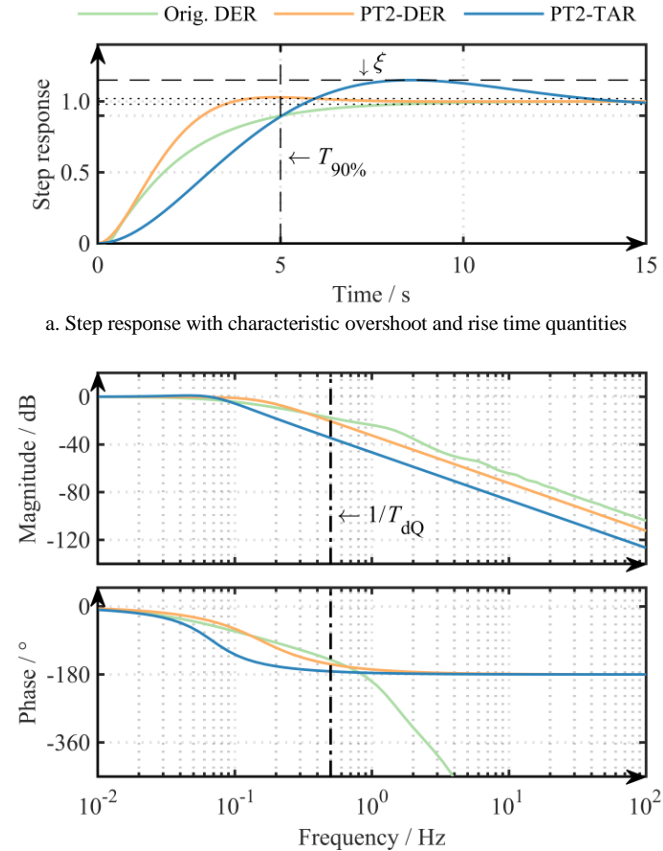

a. Step response with characteristic overshoot and rise time quantities

b. Frequency response with kink frequency of the dominant DER PT1 element

Figure 3. Comparison of step and frequency response of a detailed DER model (Orig. DER) with PT2 approximations based on generic TAR step response and detailed DER model, respectively. (DER based on WF-FRC model with $T_U = 0.02$ s, $T_{dQ} = 2$ s, $K_q = 0.5$, $T_q = 0.2$ s, $T_I = 0.1$ s, $T_g = 0.2$ s)

Subplot 3a depicts the differences in step response approximation. While the approximation based on TAR hits $T_{90\%}$ accurately, the overshoot $\xi$ is assumed to be worse compared to the real DER model. In contrast, the approximation based on detailed DER has less $T_{90\%}$ and only slight $\xi$. This is acceptable since the approximation was performed mainly in the frequency domain and the static gain is still achieved accurately and quickly. The estimation thus includes a more critical behavior.

The evaluation of the frequency response in Subplot 3b shows that PT2-DER hits the magnitude in a sufficient manner, but underestimates for higher frequencies. As a mitigation for this deficit, it can be considered that the largest time constant in the DER control loop $T_{dQ}$ significantly attenuates higher frequencies. Furthermore, it is advantages that the corresponding kink frequency is reached, before the phase of the detailed DER model reaches the critical phase -180 ° (negative feedback).

## III. STABILITY ASSESSMENT OF NONLINEAR MIMO SYSTEM

Preliminary work [13] has introduced the stability criterion from Moylan and Hill to achieve a stability statement for systems with $Q(U)$-characteristic control – conservative but without the need to know specific transfer function parameters. Furthermore, the so-called Popov criterion for evaluation of a nonlinear single-input single-output (SISO) system with given transfer function parameters was discussed as valuable alternative. However, an application to a MIMO system was not recommended due to a real value that has to be found to fulfill an evaluation statement. This can only be achieved by iterative trial and error. Therefore, this paper introduces an evaluation method for controls with known parameters based on the so-called circle criterion.

### A. Circle Criterion for the General MIMO Case

According to [24], the stability of a closed loop MIMO system of the general type presented in Figure 5 can be checked as follows: One limits the nonlinearity to the sector $\psi \in [M_\alpha, M_\beta]$. Then the closed loop MIMO system is stable if and only if

$$\boldsymbol{\Omega}(s) = (\boldsymbol{I} + \boldsymbol{M}_\beta \widetilde{\boldsymbol{G}}(s)) \cdot (\boldsymbol{I} + \boldsymbol{M}_\alpha \widetilde{\boldsymbol{G}}(s))^{-1} \tag{3}$$

is strict positive real.

### B. Application of the Circle Criterion to the MIMO System of Distributed Energy Resources With Q(U)-Characteristics

Assuming the presented DER models of section II and especially taking into account the nature of the nonlinear $Q(U)$-characteristic in Figure 4, one can conduct that the static nonlinearity is decoupled, i.e. for $i = 1, \ldots, n$ it holds that:

$$\mu_i = \psi_i(e) = \psi_i(e_i) \qquad \text{with } \psi_i \in [0, \beta_i]. \tag{4}$$

Each static nonlinearity $\psi_i$ belongs to the sector $[0, \beta_i]$, i.e., is enclosed by two lines with the slopes $\alpha_i = 0$ and $\beta_i$. Therefore, one arrives at $M_\alpha = 0$ and $M_\beta = \text{diag}(\beta_1, \ldots, \beta_n)$. Applying the circle criterion, the considered MIMO system is stable if

$$\boldsymbol{\Omega}(s) = \boldsymbol{I} + \boldsymbol{M}_\beta \widetilde{\boldsymbol{G}}(s) \tag{5}$$

is strict positive real.

This condition can be checked with the criterion of [25], cf. appendix VI.C, for given values $\beta_1, \ldots, \beta_n$. By an iterative process, the maximum possible distributions of $\beta_i$ can be found. As this method only specifies a range with guaranteed stability, no statement is made about the real stability limit.

## IV. VERIFICATION

The introduced stability criterion in section III is applied to four different high-voltage distribution grids. Subsequently, a comparison with the criterion introduced in [13] and stability bounds based on simulations with DIgSILENT *PowerFactory* 2021 is carried out. The type WF-FRC is used as DER model.

### A. Benchmark Grids

Four high-voltage distribution grids (DGs), two synthetic ones (sDG) and two realistic ones (rDG), are used for benchmarking purposes. While sDG1 was introduced in [3, 13], the development of sDG2 was based on realistic grid data from Germany within the *SimBench* project [26]. The sDG2, specified as *1-HV-mixed—0-no_sw*, was adapted in such a way that the 220-kV connection point (CP) to the higher grid level was shifted to 380-kV through exchange of the transformers, cf. Figure 6. Both rDG are located in the transmission area of *50Hertz Transmission GmbH* (TSO) in the eastern part of Germany and thereby represent different DER penetration rates. A key factor is introduced to further make these grids comparable in respect to DER penetration:

$$\rho = \frac{\sum P_{\text{DER inst}}}{l_{\text{Grid}}}. \qquad (6)$$

As a reference, the authors present the average penetration factor for Germany: $\rho_{\text{GER}} \approx 57$ kW/km. It is to mention that rural DGs usually have a higher DER penetration than urban DGs. TABLE I summarizes the key characteristics of the grids. Of course, $\rho$ is significantly lower for the given realistic DGs than for the synthetic ones, but evaluations have shown that there are also DSO grids within the TSO area where $\rho$ is above 1000 kW/km.

TABLE I. OVERVIEW OF BENCHMARK GRID CHARACTERISTICS.

| Grid | CPs to 380 kV | No. of nodes | $\sum P_{\text{DER}}$ in MW | $\rho$ in kW/km |
|---|---|---|---|---|
| sDG1 | 2 | 50 | 480 | 1400 |
| sDG2 | 3 | 61 | 1560 | 1440 |
| rDG1 | 2 | 65 | 135 | $\approx 200$[a] |
| rDG2 | 3 | 210 | 840 | $\approx 600$[a] |

a. As concrete branch length of grid is unavailable, $\rho$ of related DSO grid area is presented instead.

### B. Application to Benchmark Grids

Using the in-house developed toolbox *powerfactory-utils* [27], an automatic export from PowerFactory to an interoperable JSON based grid format can be performed conveniently. Subsequently, the computation of the nodal voltage sensitivities $K_Q$ is executed as required to build the linear transfer function matrix $\widetilde{G}(s)$ as in (1). In the following, the system is evaluated for a distribution of $Q(U)$-characteristic slopes $M_\beta$ with the presented circle criterion. In an iterative loop, the slope is now increased until the stability of the system can no longer be guaranteed based on this criterion. Further, it holds:

- each DER has the same $Q(U)$-characteristic slope $m$,
- as the special case of no dead band within the $Q(U)$-characteristic is assumed, it follows $\beta = m$.

TABLE II compares the results of the stability assessment introduced in this paper (Circle) with the method presented in the previous work [13] (Robust) as well as manual evaluation using *PowerFactory*. As evaluation measure, the recommendation given in the TAR high-voltage [14] is: $m = [6, 20]$ %/p.u.

Applying only the robust criterion to different DER representations, PT2-TAR leads to worse results, due to a higher closed loop gain that is being assumed[1]. Utilizing the circle criterion instead, a higher stability limit can be guaranteed, regardless of the DER representation. Furthermore, it can be stated that the result using the original DER transfer functions is the less conservative one. This effect follows from the conservative simplification steps. Utilizing the PT2-DER approximation, the results are significantly better than with PT2-TAR. This is not surprising as the PT2-TAR model represents the worst DER that is still allowed within the TAR specifications, where the PT2-DER tries to approximate the actual DER dynamics.

Advantageous, however, is that the circle criterion can be applied even without knowledge of concrete DER parameters using the PT2-TAR approximation. Thus, a much higher stability limit for the $Q(U)$-characteristic slope can be guaranteed in comparison to the robust criterion, e.g. see TABLE II, sDG1: 18.6 %/p.u. against 6.8 %/p.u. The advantage of PT2-DER is a higher computation speed due to the reduced model complexity within the iterative evaluation process. In addition, the evaluation with each DER model would benefit from a smart step size control, which, however, is not part of this investigation.

In *PowerFactory*, the grids are evaluated using detailed DER models of the type WF-FRC, cf. TABLE III. A broad range of disturbance variables is conceivable, but is not fully applied yet due to the high effort and the associated evaluation time. Thus, only the ramp-like increase of DER feed-in power has been performed. Here, a voltage response that showed a significant decay within 10 s after reaching the full feed-in power was considered as asymptotically stable. The results computed based on the circle criterion are higher than the recommended maximum slopes. For rDG1 even no threshold value could be determined, because too few reactive power resources were installed in the grid.

---

[1] For PT2-TAR approximation a worst case overshoot of $\xi = 15$ % is set, instead the maximum closed loop gain for a broad range of parameterization is assumed as 1 for Orig. DER and PT2-DER.

TABLE II. RECOMMENDED MAXIMUM SLOPE OF THE $Q(U)$-CHARACTERISTIC IN %/p.u. FOR GIVEN BENCHMARK GRIDS BASED ON DIFFERENT STABILITY EVALUATION METHODS AND DER REPRESENTATIONS.

| Criterion | Type of DER representation | | |
|---|---|---|---|
| | Orig. DER | PT2-DER | PT2-TAR |
| sDG1 | | | |
| Robust [13] | 6.8 | 6.8 | 5.9 |
| Circle | 55 | 29.7 | 18.6 |
| PowerFactory | 77 | – | – |
| sDG2 | | | |
| Robust [13] | 3.9 | 3.9 | 3.4 |
| Circle | 15 | 9.7 | 5.9 |
| PowerFactory | 40 | – | – |
| rDG1 | | | |
| Robust [13] | 15.7 | 15.7 | 14 |
| Circle | 131 | 72 | 45 |
| PowerFactory | No limit, because too few reactive power resources. | | |
| rDG2 | | | |
| Robust [13] | 3.3 | 3.3 | 3.0 |
| Circle | 25.3 | 13.5 | 8.8 |
| PowerFactory | 41.8 | – | – |

## V. CONCLUSION AND OUTLOOK

The authors present ongoing work on the evaluation of multi-input multi-output reactive power control systems in distribution grids. The work focusses on the interdependent $Q(U)$-characteristic control. Normative technical guidelines like [14] define the range of the recommended characteristic slope. However, a generic computation approach is not given in these guidelines, therefore the use of $Q(U)$-characteristic control beyond the given limits is the concern of the DSO. Subsequently, an improved stability assessment method based on the circle criterion is introduced. The new method enables the computation of a guaranteed stability range for the $Q(U)$-characteristic slope taking into account the parameters of the DER transfer function. The stability assessment was applied to four different high-voltage benchmark grids. The results show a wider stability range in comparison to the previous work [13]. Simulations in *PowerFactory* confirm the improved quality of the new stability evaluation approach. A big drawback of the simulations done in *PowerFactory* is that the evaluation time as well as the number of disturbance events is much higher.

Furthermore, the authors present two approaches for control loop simplifications. Both are based on PT2 approximations, the first case (PT2-DER) offers a lower computing time and the second case (PT2-TAR) allows the application even without knowledge of the control loop parameterization. As drawback to the broader applicability, the results are a bit more conservative than when using the detailed DER model.

Application use cases are the grid-wide equipment of DERs with a $Q(U)$-characteristic based voltage control in distribution grids with a high DER penetration. Furthermore, the application of a distributed $Q(U)$ fallback control for a centralized grid-wide setpoint-based voltage control is conceivable.

Future work will focus further on the grid modeling. As presented in [20], a better representation of the voltage dependent nodal sensitivities is possible but sophisticated, since it adds an additional nonlinearity. Even a more direct and maybe simplified representation of the grid admittances is applicable, cf. [28] and opens up new stability assessment approaches. In addition, it is planned to adopt other approaches, such as [29, 30], for the stability assessment of nonlinear systems. As part of a concluding benchmark, the Nyquist method is intended to be used for a linearly assumed MIMO system. Furthermore, an automatic selection of critical DER expansion scenarios from a diverse scenario pool is envisioned through implementation of the Common Rank Approximation method [31].

The source code for the derivation of the introduced simplified DER models and the application of the stability criteria will be made available along with the JSON based grid topology of benchmark grid sDG2.

## VI. APPENDIX

### A. Q(U)-Characteristic Curve

As introduced in previous work [13], a nonlinear and even asymmetric $Q(U)$-characteristic can be represented by a sector enclosing this curve between two lines with slope $\alpha$ and $\beta$, respectively. Even characteristics with variable or optimized "curved" slope $m$, as shown in [32], can thus be taken into account. The slope $m$ is defined as:

$$m := \frac{\Delta Q/P_r}{\Delta U/U_n} \qquad [m] = \frac{\%}{\text{p.u.}}. \qquad (7)$$

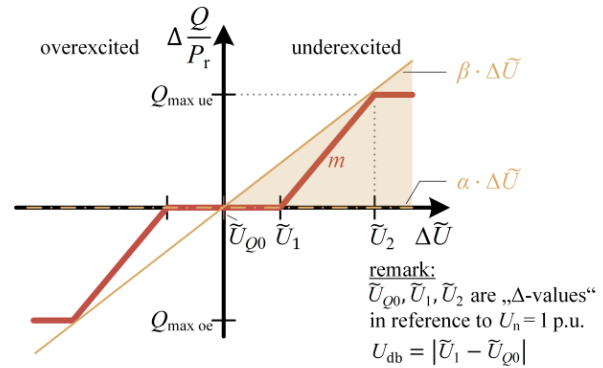

Figure 4. $Q(U)$-characteristic (red) with sector nonlinearity mapping (yellow). Note: The lower bound is identical with x-axis, hence α is zero.

### B. Model Configuration of Different DER Types

TABLE III summarizes the model configurations of different DER types and PT2 approximations from section II.B. In general, a voltage-averaging (VA) block feeds into the $Q(U)$-characteristic, which further feeds into the reactive power control loop. However, the VA can be combined with the downstream reactive power control loop, since, because of the gain of 1, the nonlinearity $\psi$ (in our case the $Q(U)$-characteristic) can be swapped with the VA [13]. Thus, it develops the summarized reactive power control loop $G(s)$. Furthermore, the migration from individual unit control to farm control is characterized by an additional communication delay. This dead time block with the time constant $T_g$ is set between the setpoint tracking $Q_R$ and unit/inverter current control $\widetilde{Q}_{set}$.

Finally, Figure 5 shows a general representation of the minimal control loop that can be derived from any DER model type. Based on Figure 2 and according to (1), $\widetilde{G}$ here subsumes the linear subsystems of the DER transfer functions $G$ and the nodal sensitivity matrix $K_Q$. The nonlinear part $\psi$ represents the $Q(U)$-characteristics. Thus, it holds:

$$\mu = \psi(-\sigma) = -\psi(\sigma),$$

where for $r \equiv 0$ Figure 5 applies.

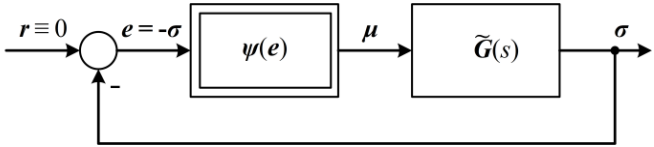

Figure 5. Minimal structure of a general nonlinear MIMO system consisting of the nonlinearity $\psi$ and the linear part $\widetilde{G}(s)$.

### C. Assessment of Strict Positive Realness of a Transfer Function Matrix

A transfer function matrix $\Omega(s)$ is strict positive real if $\Omega(s-\varepsilon)$ is positive real for any $\varepsilon > 0$ [24]. A generally valid criterion is not yet known. However, for the purpose of this paper, the result of [25] can be applied and the transfer function matrix $\Omega(s)$ is described in state space through the matrices $A$, $B$, $C$, $D$. Subsequently, according to (5),

$$\Omega(s) = I + M_\beta \widetilde{G}(s)$$
$$= C(sI - A)^{-1} B + D \quad (8)$$

is strict positive real if all eigenvalues of $A$ have negative real parts (11) and the matrix

$$N = \begin{pmatrix} -A + B\,Q^{-1}\,C & B\,Q^{-1}\,B^T \\ -C^T\,Q^{-1}\,C & A^T - C^T\,Q^{-1}\,B^T \end{pmatrix} \quad (9)$$

with

$$Q = D + D^T > 0 \quad (10)$$

has no eigenvalues on the imaginary axis (12). Thus,

$$\mathrm{Re}(\mathrm{eig}(A)) < 0, \quad (11)$$
$$\min(|\mathrm{Re}(\mathrm{eig}(N))|) > \delta \quad (12)$$

must hold. Here, $\delta$ is used as a bound for treating values as zero. As the evaluation is often to be very sensitive in this respect, $\delta$ has to be above the floating point accuracy and is set to $\delta = 10^{-8}$ in this paper. For the special case $Q = D + D^T = 0$, [25] also introduces a transformation for $A$, $B$, $C$ and $D$ such that (9) can be computed.

TABLE III. PROPOSED MODEL CONFIGURATION OF DIFFERENT DER TYPES AND RELATED PT2 APPROXIMATIONS.

| DER type | Control loop $G(s)$ | | Ref. |
|---|---|---|---|
| | Voltage averaging | Reactive power control loop | |
| (i) WF-FRC | | [block diagram: $Q_{ref} \to \frac{1}{1+sT_{dQ}} \to Q_{set} \to \Delta Q_{set} \to K_q\frac{1+sT_q}{sT_q} \to Q_R \to e^{-sT_G} \to \widetilde{Q}_{set} \to \frac{1}{1+sT_I} \to Q_{PCC}$, with $Q_{dist}$ disturbance] | [13, 18, 33] |
| | $\frac{1}{1+sT_U}$ | $T_{dQ} = 2$ s, $K_q = 0.5$, $T_q = 0.2$ s, $T_I = 0.1$ s, $T_g = 0.2$ s | |
| (ii) WF-DFIG | $T_U = 0.02$ s | [block diagram: $Q_{ref} \to \Delta Q_{ref} \to K_q\frac{1+sT_q}{sT_q} \to Q_{set} \to \frac{1+sT_{ft}}{1+sT_{fv}} \to Q_R \to e^{-sT_G} \to \widetilde{Q}_{set} \to$ Local U-control $\to \frac{1}{1+sT_I} \to Q_{PCC}$, with $Q_{dist}$] | [17, 18, 33] |
| (iii) PVF | $\frac{1}{(1+sT_U)^3}$ | A PV inverter with fast power control $T_I$ according to [34] can be extended by a farm control. The emerging control loop can be established analogously to (i). | [9, 34] |
| | $T_U \approx 0.004$ s | $T_{dQ} = 2$ s, $K_q = 0.5$, $T_q = 0.2$ s, $T_I = 0.0033$ s, $T_g = 0.1$ s | |
| (iv) PT2-TAR | Not explicit depictable. | [block diagram: $Q_{ref} \to \frac{\kappa}{1+s\,2DT+s^2T^2} \to Q_{PCC}$, with $Q_{dist}$] Based on the generic step response given in TAR [14]. $\kappa = 1$, $D = 0.517$, $T = 2.335$ s | |
| (v) PT2-DER | | Based on the frequency response of detailed DER model (i). $\kappa = 1$, $D = 0.747$, $T = 1.028$ s | |

## D. SimBench High-Voltage Grid

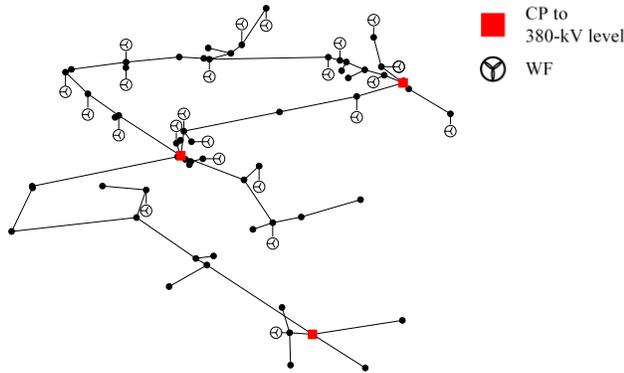

Figure 6. High-voltage distribution grid sDG2: *1-HV-mixed—0-no_sw*.

ACKNOWLEDGMENT

This paper was funded within the project STABEEL by the Deutsche Forschungsgemeinschaft (DFG, German Research Foundation) – 442893506.